# The Role of Chinese-American Scientists in China-US Scientific Collaboration: A Study in Nanotechnology

Xianwen Wang*[1,2], Shenmeng Xu[1,2], Di Liu[1,2], Yongxia Liang[3]

1. WISE Lab, Dalian University of Technology, Dalian 116085, China
2. School of Public Administration and Law, Dalian University of Technology, Dalian 116085, China
3. National Science Library, Chinese Academy of Sciences, Beijing 100190, China

* Corresponding author.
Email address: xianwenwang@dlut.edu.cn
Website: xianwenwang.com

**Abstract** In this paper, we use bibliometric methods and social network analysis to analyze the pattern of China-US scientific collaboration on individual level in nanotechnology. Results show that Chinese-American scientists have been playing an important role in China-US scientific collaboration. We find that China-US collaboration in nanotechnology mainly occurs between Chinese and Chinese-American scientists. In the co-authorship network, Chinese-American scientists tend to have higher betweenness centrality. Moreover, the series of polices implemented by the Chinese government to recruit oversea experts seems to contribute a lot to China-US scientific collaboration.

## 1. Introduction

In the long history of scientific collaboration, the earliest documented collaborative scientific paper was published in 1665, which was attributed to Hooke, Oldenburg, Cassini, and Boyle (Beaver & Rosen, 1978). It was not until the middle of the 18th century that the growth of scientific collaboration increased dramatically (Luukkonen, Persson, & Sivertsen, 1992).

Nowadays, scientific collaboration has become a very important pattern of scientific research. This phenomenon can be measured by co-authorship of published papers. According to Kostoff's study, in terms of research articles, especially in cutting-edge technologies, such as nanotechnology and energetic materials, China has grown significantly and is among the leaders in the world. Moreover, it has been shown that there was a substantial increase in highly cited documents when foreign collaborators, especially from the USA, were included (Kostoff et al., 2007).

The motivation of international scientific collaboration is complicated. Countries benefit from international collaborations in greater visibility and higher citation impact (Glänzel & De Lange, 2002). In addition, the demand for international collaboration is much stronger in countries with low scientific production than in advanced countries (Davidson Frame & Carpenter, 1979). Scientific collaborations of China were analyzed both on country level (He, 2009; Jin & Rousseau, 2005) and institution level (Tang & Shapira, 2011).

On country level, Jin & Rousseau (2005) observed the exponential growth of internationally co-authored papers of China. Furthermore, Tianwei He's results indicate that international collaboration publication output between China and the G7 countries has grown exponential thanks to the growth of science in China, and notably, the USA is the most important collaborative country for China (He, 2009).

On institution level, Li Tang and Philip Shapira's research focused on the China–US scientific collaboration in nanotechnology. Through the collaboration analysis of institutions, they concluded that "The pattern of China's nanotechnology R&D collaboration with the US is asymmetrical, with a relatively small number of elite Chinese research organizations and universities working with a wide array of US universities" (Tang & Shapira, 2011). However, this is worth discussing, because we need to consider the fact that the biggest institution in China, namely the Chinese Academy of Sciences (CAS), has over 50, 000 researchers, which is much more than any other university in China. In other words, CAS and a few top universities have dominated scientific research in most fields in China. As a result, we consider that collaboration analysis of institutions is not sufficient to reveal the patterns of China-US scientific collaboration.

Every two or more researchers is the fundamental unit of collaboration, because at the most basic level, it is people who collaborate, not institutions. Inter-institutional and international collaboration need not necessarily involve collaboration between every two or more individuals, but it is the individuals that play an important role in the beginning of a collaboration (Katz & Martin, 1997). Melin (2000) suggested the collaborations are characterized by a high degree of self-organization on individual level. Moreover, regarding collaboration cosmopolitanism, Bozeman & Corley (2004) found that most researchers tend to work with people in their own work group and people within relatively short geographical distance.

In our research, different from the existing analyses at the level of institutions, we go deep into the collaboration of individual scientists between China and the USA, and especially focus on the role of Chinese-American scientists in China-US scientific collaboration.

## 2. Data and Methods

### 2.1 Data sources

In recent years, investing immense human and financial resources in nanotechnology, China has achieved great development in this field. According to related research, China has become the second largest global producer of nanotechnology papers (Kostoff, Koytcheff, & Lau, 2007; Tang & Shapira, 2011). Furthermore, the work of Chen shows that China has overtaken the USA in nanotechnology research paper production in the 2008/2009 time frame (Chen et al., 2009).

In this paper, we select nanotechnology as the research object. Our data is collected from the SCI-Expanded citation database in *Web of Science*. We mainly introduce our data collecting methods for collaboration network analysis (Part 3.2), because the process is complicated and Part 3.2 is the main analysis of our paper.

We search the data at the same time in 3 ways, which are title search, topic search and journal search. All the search keywords are referenced from Meyer, Debackere, & Glänzel (2010).

Based on the fact that keywords starting with nano are considered as core words about nanotechnology, the topic search string is:

*TS=(nano* not (nano2 or nano3 or nano4 or nano5 or nano[-]secon* or nano[-]gram* or*

*wnanomol* or nanophtalm* or nanogeterotroph* or nanomeli* or nanoplankton* or nanokelvin* or nanocurie* or wnano[-]curie* or nanos or anos1 or nanoproto* or nanophyto* or nanoflagel* or nanobacter* or nano[-]bacter* or nanospray* or nano[-]spray*))*

Other keywords are considered as peripheral core words about nanotechnology, which are used only in title search. The search string is edited as:

*TI=(quantum[-]dot* or quantum[-]wire* or carbon[-]tub* or carbontub* or buckytub* or bucky[-]tub* or fullerene[-]tub* or self[-]assembled[-]monolayer* or self[-]assembl*[-]dot* or single[-]electron[a-z]tunnel* or single[-]molecul* or molecul*[-]motor* or molecul*[-]ruler* or molecul*[-]wir* or molecul*[-]devic* or molecular[-]engineering* or molecular[-]electronic* or bionano* or nanomet*[-]chip* or nanomet*[-]layer* or nanomet*[-]diamet* or nanomet*[-]electron* or nm[-]engin* or nm[-]chip* or nm[-]layer* or nm[-]diamet* or nm[-]electron* or submicro*[-]engin* or submicro*[-]chip* or submicro*[-]layer* or submicro*[-]diamet* or submicro*[-]electron* or molecul*[-]beacon* or molecul*[-]engin* or molecul*[-]manufact* or biochip* or dna[-]cmos* or fulleren[-]pip* or molecul*[-]self[-]assembl* or self[-]assembl*[-]multilayer*)*

Meanwhile, 22 journals are considered core journals about nanotechnology, and all the papers published in these journals are collected. We set the search string as:

*SO=(ACS Nano or Current Nanoscience or Fullerenes Nanotubes and Carbon Nanostructures or IEE Proceedings-Nanobiotechnology or IEEE Transactions on Nanobioscience or IEEE Transactions on Nanotechnology or IET Nanobiotechnology or International Journal of Nanomedicine or Journal of Computational and Theoretical Nanoscience or Journal of Experimental Nanoscience or Journal of Nanoparticle Research or Journal of Nanoscience and Nanotechnology or Lab on a Chip or Nano or Nano Letters or Nano Today or Nanoscale Research Letters or Nanotechnology or Nature Nanotechnology or Nanobiology or Nanostructured Materials or Journal of Nanophotonics)*

At the end, the 3 query ways mentioned above are combined to get the final data for our research.

The time span is set as from 2007 to 2010, because it is only from 2007 onwards that the link between authors and addresses is registered in *Web of Science*. Another important reason for this is that *Web of Science* didn't provide authors' full name until 2007. Without the full names, it is very difficult for us to distinguish the names of each author. Especially in China, some first names, for example, Wang, Zhang, Li, are very common.

Moreover, the addresses including Hong Kong are also excluded. As a special administrative region of China, Hong Kong is much more international than mainland China, which would cause large deviation to the result.

Papers including two or more addresses of one scientist do not denote collaboration between institutions in our analysis. When collecting data, one important rule is that we only adopt scientists' sole fulltime positions, mostly their tenure positions of professor or associate professor, in the cases of scientists who hold dual positions at both a Chinese and a US institution. In other words, by searching for the scientists' information one by one, we make sure that there is only one address per scientist that counts. We exclude the remaining papers in which one (or more) author(s) mention two or more addresses in our study.

## 2.2 Data processing

We need to normalize our data before analyzing them. In this part, we mainly deal with the processing of names.

**(1) Abbreviation names**

Because some journals don't provide full names, there are still a few abbreviation names in the data. For example, Wang J may be the abbreviation of Wang Jing or Wang Jun. So, these abbreviation names should be transformed to full names. We search the abbreviation names from their institutes and combine the information of their co-authors, to find out their full names.

**(2) Namesake**

For those authors with short given names and some common first names, for example, Wang, Zhang, Li, namesake is also a common phenomenon. For example, there are 8 authors with the same name of Wang Jing. These 8 authors come from Peking University, Nanjing University, Sichuan University, Chinese Academy of Sciences, Soochow University, SunYatsen University, University of Michigan, Royal Institute of Technology. To distinguish these authors, we add a suffix of his/her institute after the name. For example, Wang Jing-CAS indicates Wang Jing from Chinese Academy of Sciences.

### 2.3 Method

Social network refers to groups of people each one of which has connections of some kind to some or all of the others (Newman, 2001). Co-authorship network is one kind of social network, through which scientists form the network of one specific scientific field. In the network, each node represents one scientist, and two scientists are connected if they have coauthored a paper. To some degree, the network may reflect the importance and position of scientists in the scientific territory. The structure of collaboration networks "turns out to reveal many interesting features of academic communities" (Newman, 2004).

In this research, the main method adopted is social network analysis (SNA), including the network structure and indicators, for example, betweenness centrality.

## 3. Results

### 3.1 Statistical analysis

In this section, we perform statistical analysis about China's international collaboration counts and China-US collaboration counts from 2001 to 2010. Because *institution* information is not available in the pre 2007 papers, data in this part are retrieved from *Web of Science* directly according to the *Country/Territory* information. And papers in which one (or more) author(s) mention two or more addresses are not excluded. Therefore, this part only reveals a general background of China-US collaboration in this decade. However, it helps to better understand the further analysis in the *Collaboration network analysis* part.

#### 3.1.1 International collaboration and China-US collaboration

We search the number of SCI papers of China with the search string *CU= China*. And then, we use the *refine result* function of *Web of Science* to get the number of SCI papers of China collaborating with foreign countries. Meanwhile, we query with the search string *CU= (China and USA)* to get the China-US collaborative SCI papers. We collect all the data for the time span from 2001 to 2010.

Fig. 1 shows the collaboration counts of SCI papers between China and foreign countries. Over the past decade, the number of SCI papers of China keeps a relatively stable growth from 36, 497 in 2001 to 144, 243 in 2010, with a growth rate of 295%. Simultaneity, the internationally co-authored papers of China increases from 8, 633 in 2001 to 34, 634 in 2010, and the growth rate reaches 301%. The proportion of collaborative SCI papers maintains at about 24%.

The number of China-US collaborative SCI papers has a large presence. In 2001, the number of China-US collaborative SCI papers is only 3160, accounting for about 8.66% of the total SCI papers of China, and 36.60% of total international collaborative papers of China; in 2010, it has increased to 15,359, when the ratio increased to 10.64% of total and 44.35% of total collaboration. The overall trend in China-US collaboration is stable growth. The growth of China-foreign collaboration and China-US collaboration can be fitted by exponential growth.

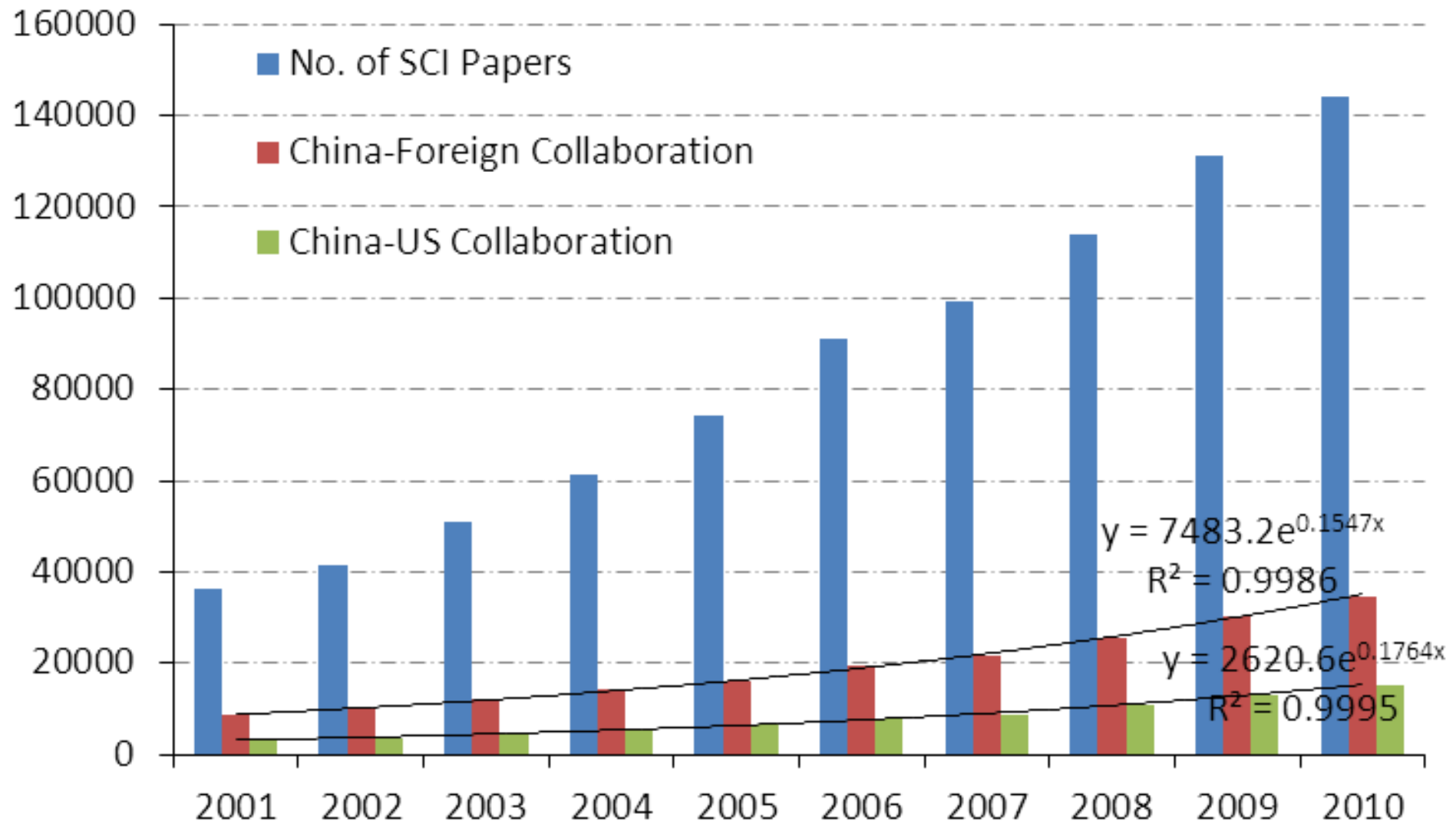

**Fig. 1** Collaboration of SCI papers among China and foreign countries

Fig. 2 illustrates the ratio of China-US collaboration in China-foreign collaboration more clearly. As is shown, from 2001 to 2010, the proportion is in a growth trend overall. By 2010, nearly half of the internationally co-authored papers of China are co-authored with the United States.

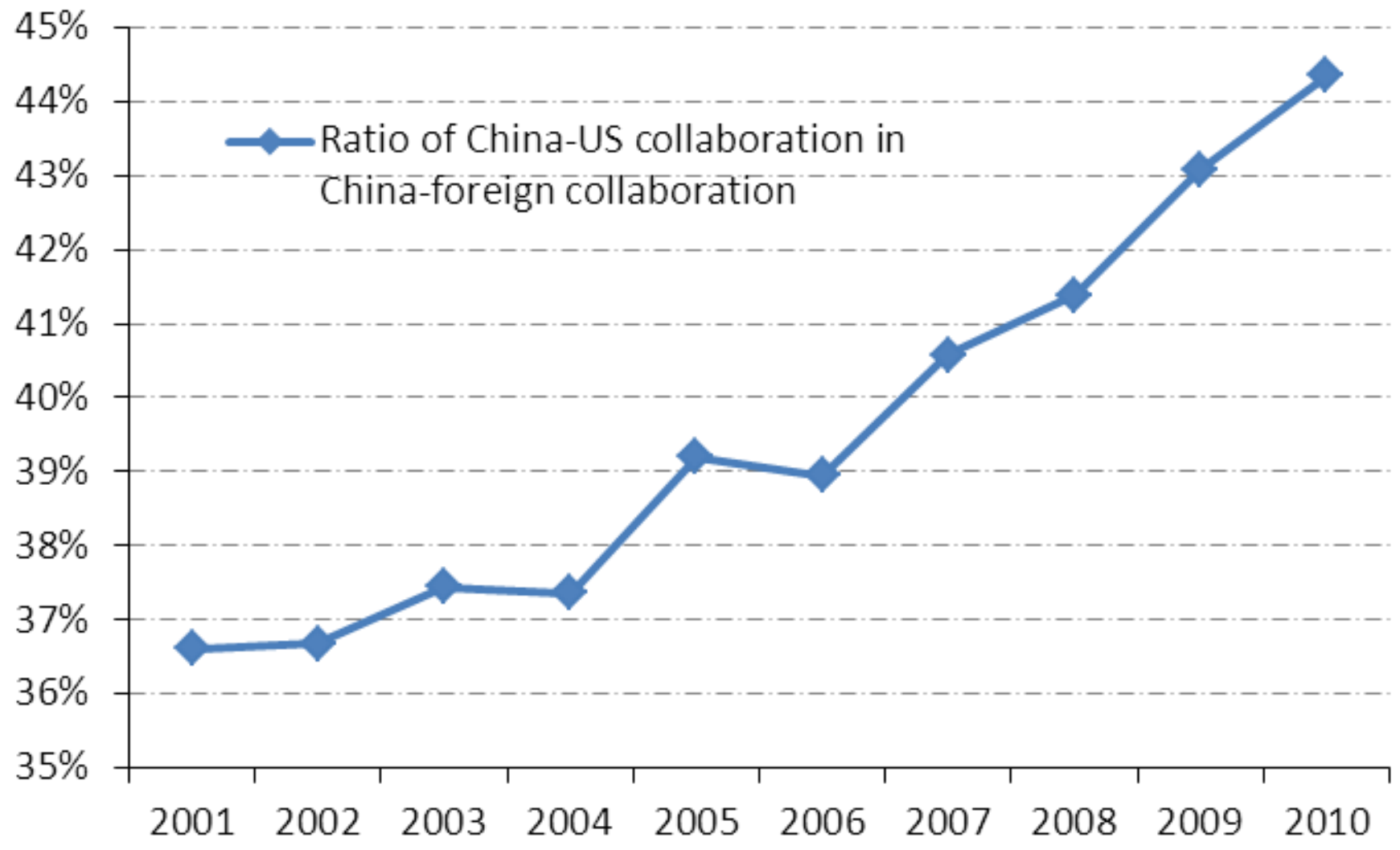

**Fig. 2** Ratio of collaboration with the USA

### 3.1.2 China-US collaborated papers in nanotechnology

**(1) The trend for the publication and co-authored counts**

Fig. 3 shows both China's SCI paper numbers and China-US collaboration trend in nanotechnology from 2001 to 2010. We can see that the number of SCI papers of nanotechnology increases fast yearly. In 2001, the number is 1934, while in 2010, it has risen to 17,516. As for China-US collaboration, during the last decade, the number of co-authored papers also keeps a steady growth, rising from 108 in 2001 to 1415 in 2010. When taken into comparison, in 2001, about 5.58% of China's nanotechnology papers are co-authored with the USA, while in 2010, the ratio is 8.08%. Although in 2004, the ratio slipped to 5.15%, after then, it keeps growing fast.

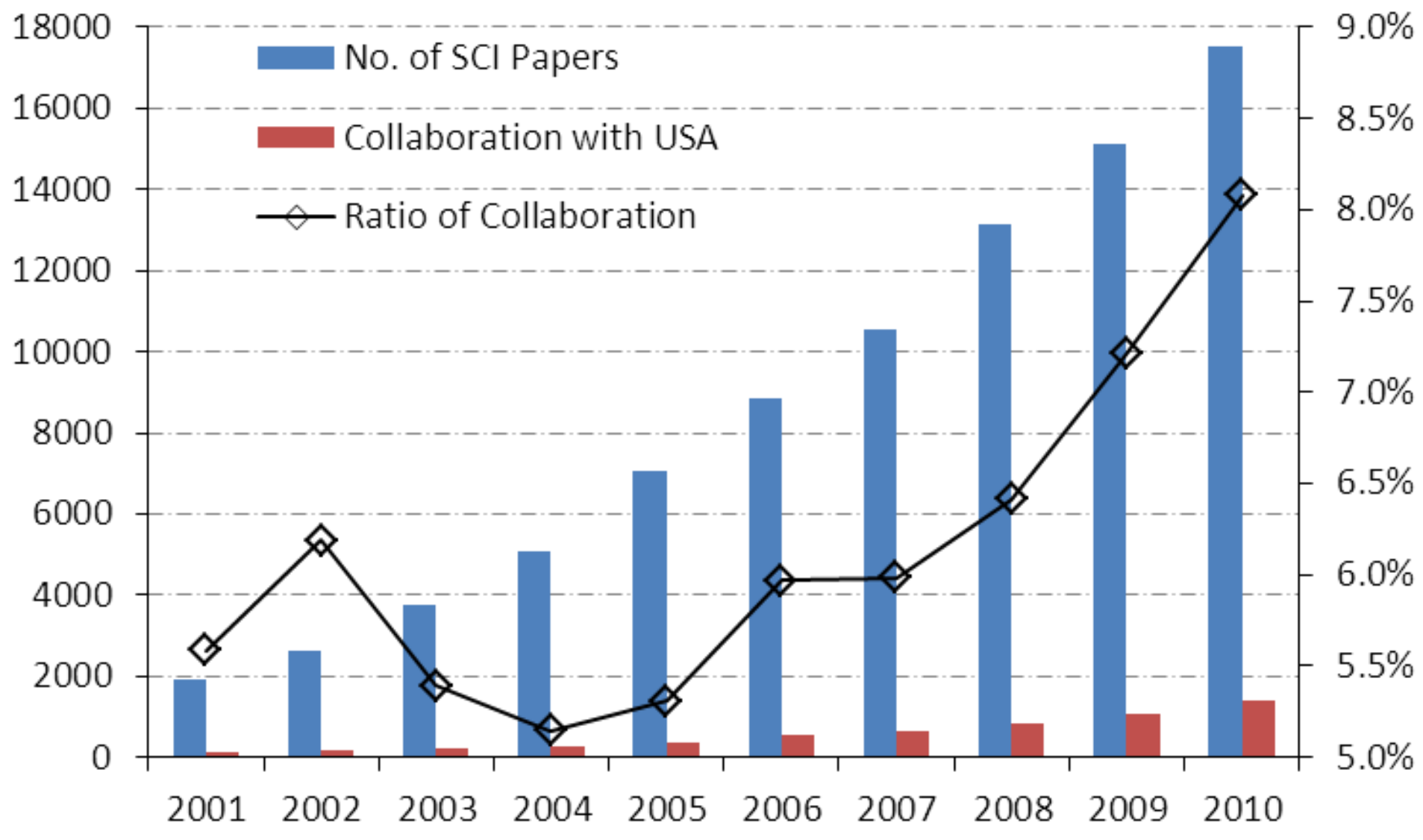

**Fig. 3** Collaboration between China and the USA in nanotechnology

**(2) Statistical analysis of authors**

Fig. 4 is the statistical analysis of the authors. There are 141 unique authors with 10 or more papers in the dataset. By confirming their addresses one by one, we find that 56 authors are from China, accounting for 39.7% of the total, while 83 are from the USA, accounting for 58.9%. The two remaining authors are from Japan and Singapore respectively. Among the 83 authors from the USA, we find that 69 are Chinese-American, which means that they are born in mainland China, received their B.S. in China, and are currently working full-time in the United States after they got their Ph.D. or did postdoctoral work in the United States. China-US scientific collaboration mainly occurs between Chinese and Chinese-American, with collaboration between Chinese and non-Chinese-American scientists representing a small proportion only (16.87%). Hence it can be concluded that Chinese-American scientists play a very important role in China-US scientific collaboration. Furthermore, we also find out that among these 69 Chinese-American authors, 30 are engaged as chair/adjunct/guest professors in China's institutions, accounting for 43.5% of the total number.

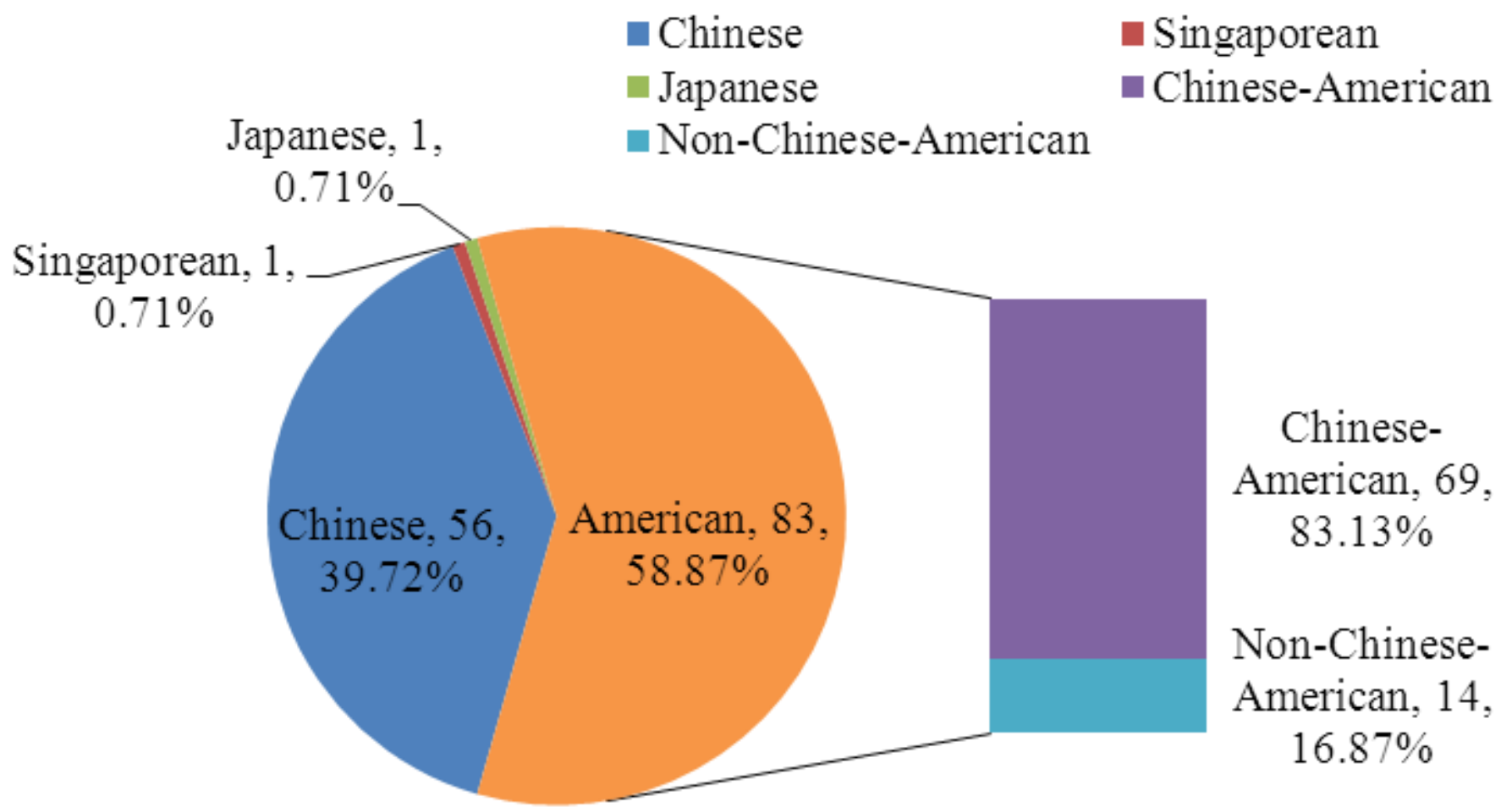

**Fig. 4** Author statistical analysis

### 3.2 Collaboration network analysis

Collaboration network analysis is one kind of social network analysis. A social network is a network of social relations, reflecting a relationship between actors. Actors in the network manifest as nodes and the relationships between them manifest as the links between the nodes. Here we focus on the relationship between the authors, in order to find the collaboration patterns between Chinese and American scientists.

608 authors who have published more than 4 papers are selected to construct the collaboration matrix using Bibexcel (http://www8.umu.se/inforsk/Bibexcel/). Then the matrix is imported to Netdraw to perform network analysis (Borgatti, 2002).

To eliminate co-authorships that resulted from occasionality, we first set the threshold to 1, which means only those authors who have more than 1 co-authored paper are kept in the network.

As is shown in Fig. 5, black nodes represent American authors, while orange nodes represent Chinese authors. The size of nodes indicates the *betweenness centrality* of authors in the network. *Betweenness centrality* often measures the nodes' ability as the media, that is, the ability of occupying the other two nodes on the shortest path. The greater the *betweenness centrality* is in the network, the more important position it occupies and the more dominant role it plays.

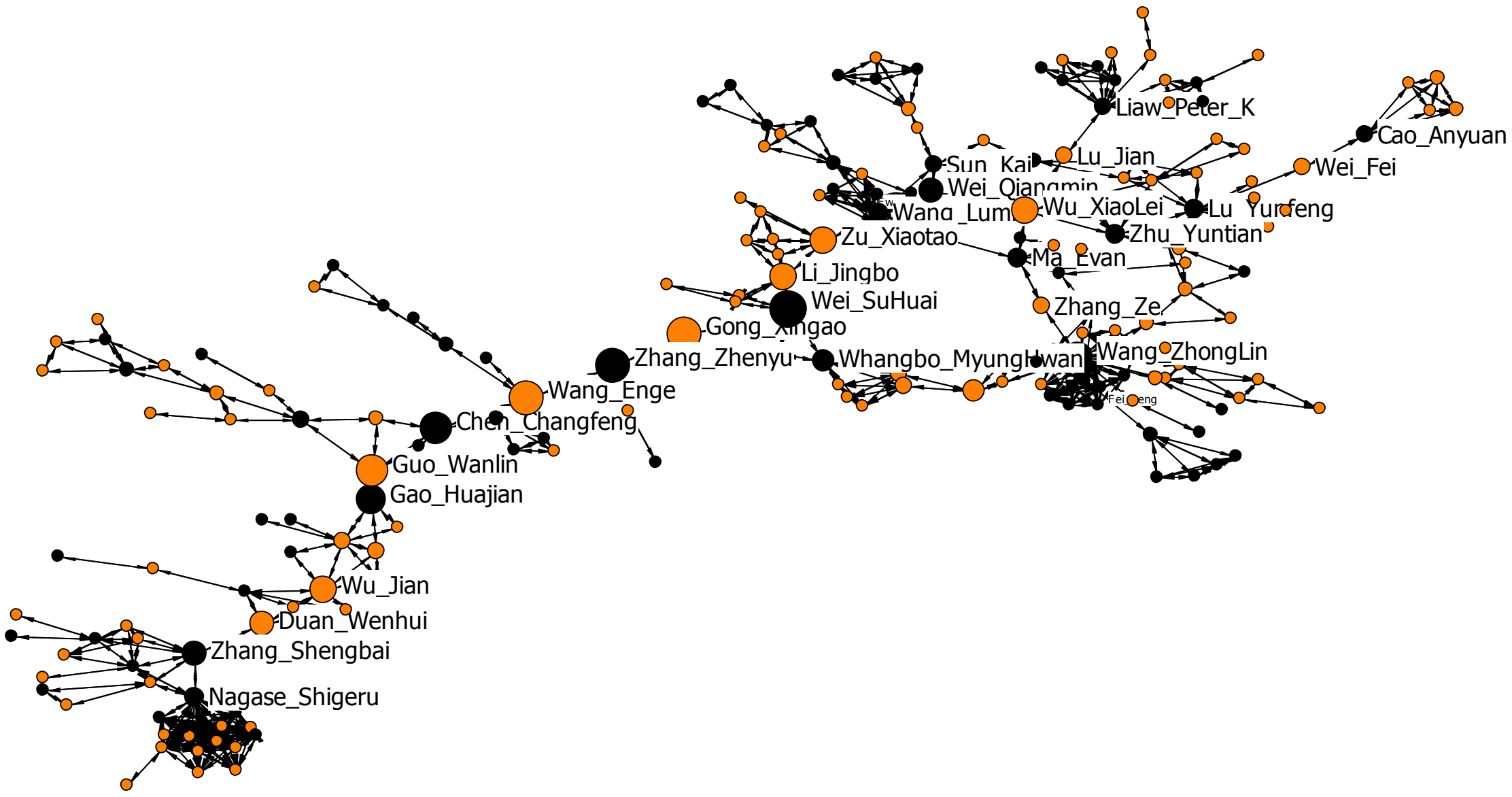

**Fig. 5** The main component of China-US collaboration in nanotechnology (cutoff is 1)

Table 1 shows the betweenness centrality of nodes in the main component. There are 204 nodes in the network, among which 93 authors are from the USA and 111 are from China.

**Table 1** Betweenness centrality of nodes in the network of Fig. 5

| | The USA | | China | |
|---|---|---|---|---|
| | No. of Nodes | Ratio | No. of Nodes | Ratio |
| Total | 93 | 100% | 111 | 100% |
| Betweenness centrality > 100 | 39 | 42% | 42 | 38% |
| Betweenness centrality > 1000 | 23 | 25% | 18 | 16% |
| Betweenness centrality > 10000 | 2 | 2% | 2 | 2% |

We divide the betweenness centrality into 3 levels, which are >100, >1000 and >10000. There are 81 authors with betweenness centrality greater than 100, among whom 39 are from the USA and 42 are from China. The number of Chinese authors is a little larger than American authors. However, for the 41 authors with betweenness centrality greater than 1000, the USA has 23, while China has 18. Furthermore, at the level of 1000 for the betweenness centrality, there are only 4 authors qualified, 2 from the USA and the other 2 from China. Generally speaking, Chinese authors and American authors hold almost equal shares for the total numbers and important authors in the network.

Then we sum up the betweenness centrality of Chinese authors and American authors separately. For the 93 American authors, the overall betweenness centrality is 108661, which is significantly greater than the total number of 96632 for the 111 Chinese authors. And, we calculate the average betweenness centrality for the two groups. The quotient of 108661 divided by 93 authors is 1168.4, which is the average centrality of American authors, while the average centrality for Chinese authors is 870.6. As a result, we can conclude that American authors have greater importance than Chinese authors in the collaboration network in general.

In Fig. 5, Wei Suhuai has the greatest betweenness centrality, which is as high as 11794. Wei Suhuai is from National Renewable Energy Laboratory (NREL). He received his B.S. from Fudan University, Shanghai, China in 1981, and his Ph.D. from the College of William and Mary in 1985. After graduation, he joined NREL in 1985. And in 2007, he was engaged as chair professor in Fudan University, Shanghai, China, which is supported by "Cheung Kong Scholars Programme of China", one of the most important plans to attract overseas talents to return to China.

Zhang Zhenyu is the other American author with highest level betweenness centrality, who is from Oak Ridge National Laboratory (ORNL). He received his B.S. in Wuhan University, Wuhan, China in 1982, and Ph.D. from Rutgers University in 1989. After that, he worked in UC Santa Barbara, University of Tennessee, Knoxville, ORNL, etc. In 2008, he accepted an offer of the "1000 Plan Recruitment Program of Global Experts" from the General Office of the Central Committee of the Chinese Communist Party, which is regarded as the highest level of program to recruit overseas Chinese talents. In January 2011, Zhang Zhenyu accepted a full time offer to work in University of Science and Technology of China (USTC).More information about these American authors is presented in Table 2. There are 22 American authors with betweenness centrality greater than 1000. Among them, 20 received their B.S. in mainland China, and 17 got their Ph.D. in the USA. The other 3 scientists got their Ph.D. in China, and after graduation, they firstly worked as post doctors in the USA, then stayed or moved to other American institutions. Among the 22 American authors, 12 are engaged as chair/guest/adjunct professor in Chinese universities, 6 for chair professor of “Cheung Kong Scholars of China” and 2 for “1000 Plan”.

**Table 2** Education experience of American authors

| | Work | B.S. | M.S. | Ph.D. |
|---|---|---|---|---|
| Zhang Shengbai | Rensselaer Polytech Inst | Jilin U, CN | UC Berkeley | UC Berkeley |
| Huang Yonggang* | Northwestern U | Peking U, CN | Harvard U | Harvard U |
| Gao Huajian* | Brown U | Xian Jiaotong U, CN | Harvard U | Harvard U |
| Chen Changfeng | U Nevada | N/A, CN | N/A, CN | Peking U, CN |
| Zhang Zhenyu* | Oak Ridge Natl Lab | Wuhan U, CN | | Rutgers U |
| Wei Suhuai* | Natl Renewable Ener Lab | Fudan U, CN | | Coll William & Mary |
| Whangbo MyungHwan | N Carolina State U | Seoul Natl U, KR | Seoul Natl U, KR | Queen's U, CA |
| Wang Zhonglin* | Gatech | Xidian U, CN | | Arizona State U |
| Song Jinhui | Gatech | Nankai U, CN | Gatech | Gatech |
| Zhang Fan | Penn State U | | | |
| Ma Evan* | Johns Hopkins U | Tsinghua U, CN | | Caltech |
| Wang Lumin* | U Michigan | Beijing Polytech U, CN | U Wisconsin Madison | U Wisconsin Madison |
| Wei Qiangmin | U Michigan | Wuhan U, CN | State U Nebraska | U Michigan |
| Liu Guokui | Argonne Natl Lab | Wuhan U, CN | Montana State U | Montana State U |
| Ren Zhifeng* | Boston Coll | Xihua U, CN | Huazhong U S&T, CN | Chinese Academy Sciences, CN |
| Sun Kai | U Michigan | Dalian U Tech, CN | Dalian U Tech, CN | Dalian U Tech, CN |
| Zhu Yuntian | N Carolina State U | Hefei U Tech, CN | Oregon Grad I S&T | U Texas at Austin |
| Cao Anyuan* | U Hawaii Manoa | Tsinghua U, CN | | Tsinghua U |
| Liaw Peter K | U Tennessee | Natl Tsing Hua U, TW | | Northwestern U |
| Lu Yunfeng* | UC Los Angeles | Jilin U, CN | Chinese Acad Sci, CN | U New Mexico |
| Wei Bingqing* | U Delaware | Tsinghua U, CN | Tsinghua U, CN | Tsinghua U |
| Zeng Xiaocheng* | U Nebraska | Peking U, CN | | Ohio State U |

* The author is engaged as chair/guest/adjunct professor in Chinese universities.

Then, we increase the threshold from 1 to 2, which means only those authors who have more than 2 co-authored papers could be kept in the network, in order to better eliminate the namesake phenomenon, and visualize the network structure more clearly.

Fig. 6 is the main component of the co-authorship nework. In Fig. 6, most authors with high betweenness centrality are from the USA.

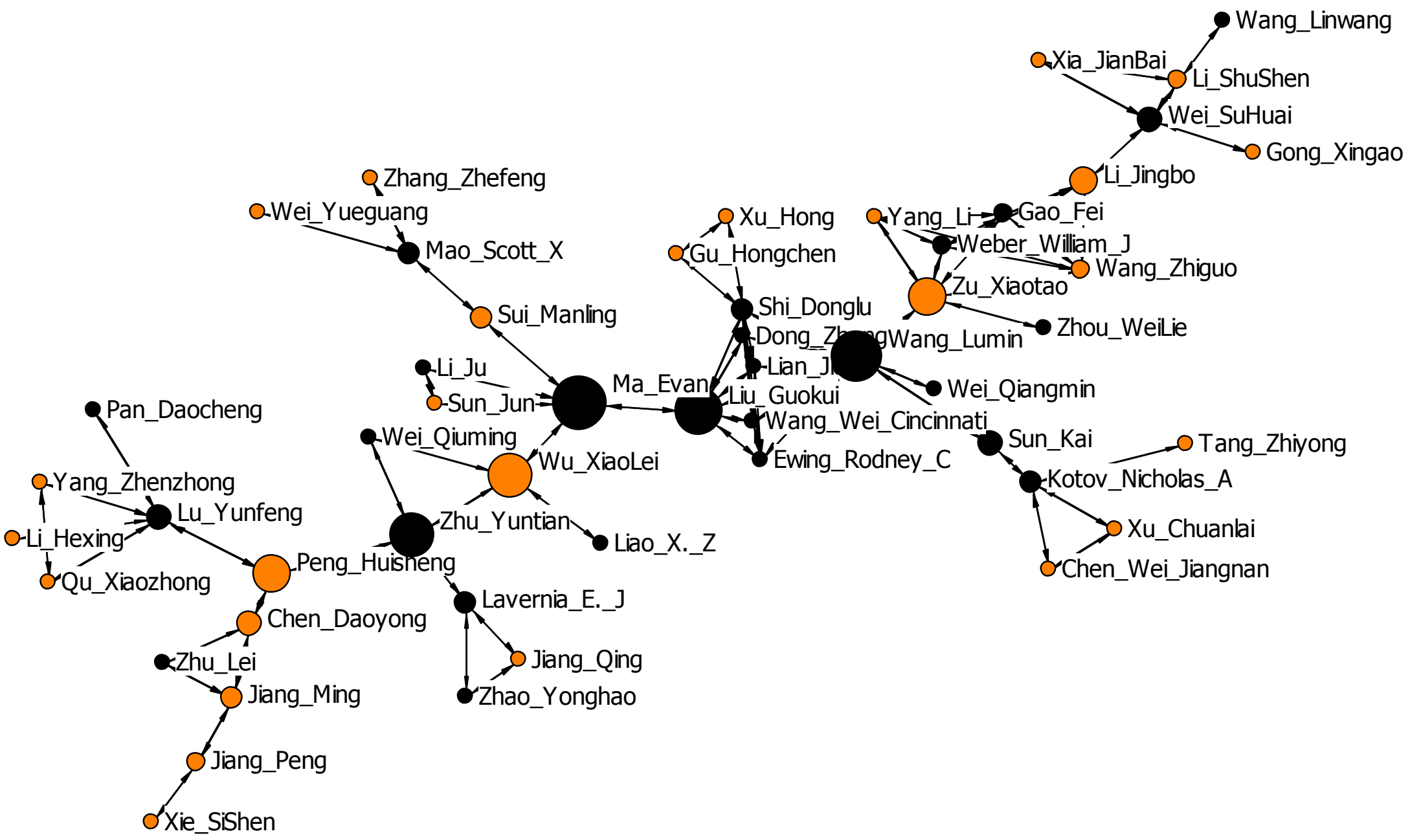

**Fig. 6** The main component of China-US collaboration in nanotechnology (cutoff is 2)

As Table 3 shows, there are 52 authors in the network in Fig. 6, among whom 26 are from the USA and 26 are from China. 23 authors have betweenness centrality greater than 100, among whom 13 are from the USA and 10 are from China. 14 authors have betweenness centrality greater than 1000, and among them 8 are from the USA and 6 are from China. However, among the 5 authors who have betweenness centrality greater than 10000, 4 are from the USA and only 1 is from China. So, we can conclude that in the network of Fig. 6, when we increase the threshold to 2, American authors play a more dominant role in the co-authorship network.

**Table 3** Betweenness centrality of nodes in the network of Fig. 6

| | The USA | | China | |
|---|---|---|---|---|
| | No. of Nodes | Ratio | No. of Nodes | Ratio |
| Total | 26 | 100% | 26 | 100% |
| Betweenness centrality > 100 | 13 | 50% | 10 | 38% |
| Betweenness centrality > 1000 | 8 | 31% | 6 | 23% |
| Betweenness centrality > 10000 | 4 | 15% | 1 | 4% |

### 3.3 International collaboration and citation

To study further on the impact of the collaboration between China and the USA, we analyze the highly-cited papers in nanotechnology of China published in 2009. Here we consider the papers cited more than 10 times as highly-cited ones, and get 125 papers in all. As is shown in Table 4, there are 72 papers authored by scientists from China only. It's worth noting that 53 papers are authored by China and foreign collaborators, occupying 42.4% of the total numbers. And among these, more than half (29 papers) are collaborated by China and US collaborators, indicating the significant impact of the foreign collaborations, especially China-US collaborations, on the increase in cited papers for China in nanotechnology.

**Table 4** International collaboration and highly cited papers

| | Number | Percentage |
|---|---|---|
| Highly cited papers (>=10 times) | 125 | 100% |
| China only | 72 | 57.60% |
| China and foreign collaborators | 53 | 42.40% |
| China and US collaborators | 29 | 23.20% |

In addition, among all the 15116 papers of China published in nanotechnology in 2009, 12295 are authored only by Chinese scientists. As Fig. 7 shows, the average citation of these papers is 5.96, relatively low compared to the papers in collaboration with foreign countries. We can see that the number of papers in international collaboration is only 2821, but they receive an average of 9.15 citations. Moreover, the average citation of the 1088 China-US collaborated papers reaches 10.75. This result is in line with the result of Kostoff's study (Kostoff *et al*., 2007), who found that collaboration with foreign countries produces a substantial increase in numbers of Chinese articles published in journals with a very high JCR Impact Factor.

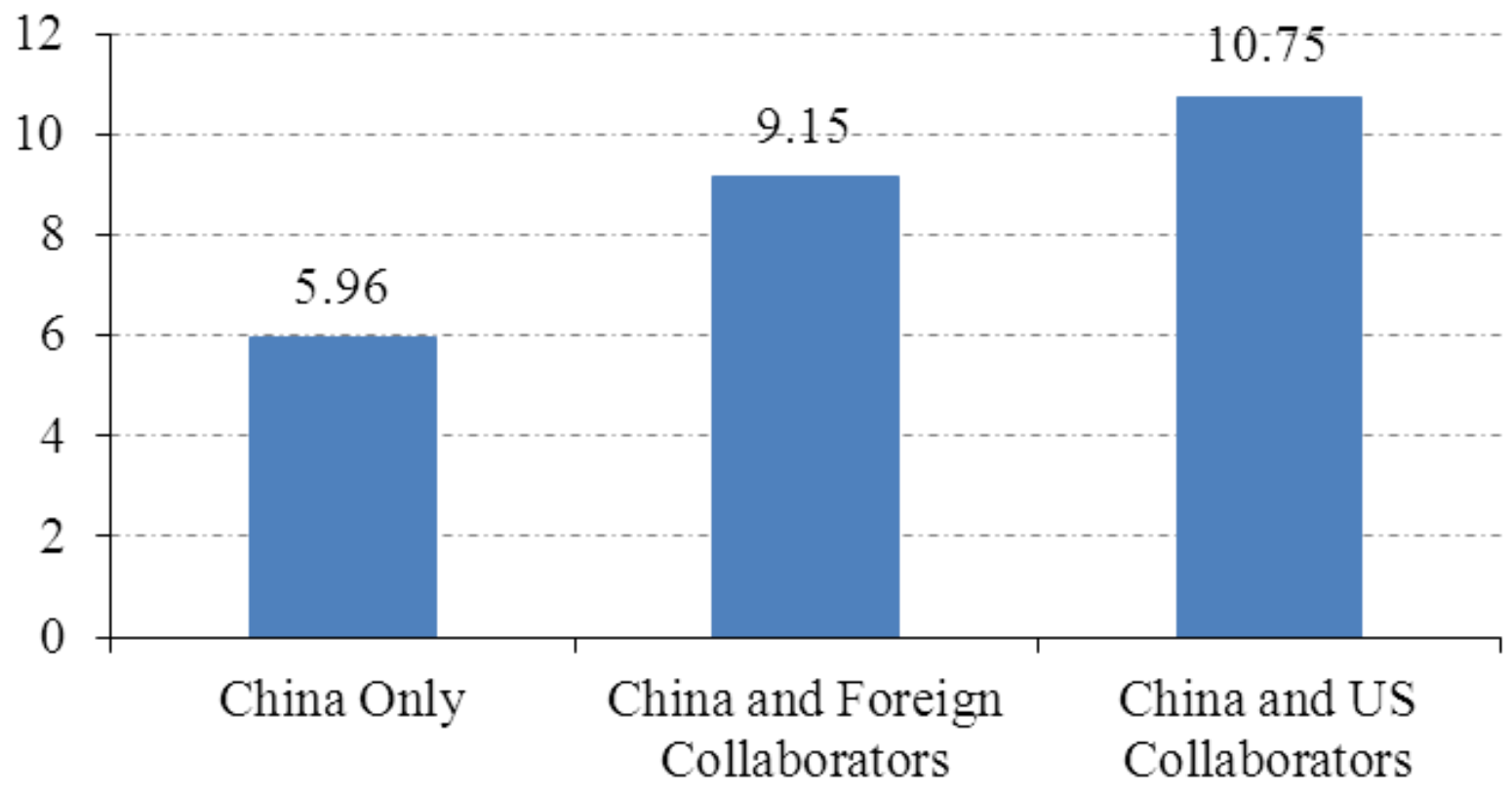

**Fig.7** International collaboration and average number of citations

## 4. Discussion

Through further inspection and analysis of China-US scientific collaboration in nanotechnology, we find that:

(1) China-US collaboration is asymmetrical. The statistical analysis shows that among the 141 authors who have published more than 9 papers, only 39.7% are from China, while 58.9% are from the USA, which means a relatively small number of Chinese scientists working with a wide array of US scientists. It is similar to the finding of the collaboration analysis at the level of institutions in Tang's research (Tang & Shapira, 2011).

(2) Most US co-authors of Chinese scientists are Chinese-American, who have received their B.S. in Chinese institutions and got Ph.D. or have postdoctoral experience in US institutions. Most US co-authors have a tenure positions of professor or Associate professor, while they still keep in touch with Chinese peers. Many of them are engaged as chair/guest/adjunct professors in Chinese institutions, which is the most significant cause to generate China-US scientific collaboration. Actually, in order to recruit the foreign talents, especially the foreign talents of Chinese origin, a series of programs are laid down, such as "1000 Plan Recruitment Program of Global Experts" from the General Office of the Central Committee of the Chinese Communist Party, "Cheung

Kong Scholars of China” by Ministry of Education of China, “100-Talents Scheme” by Chinese Academy of Sciences, etc. The remuneration packages are very attractive. For example, for the “1000 Plan Recruitment Program of Global Experts”, the one-time subsidies for each returnee is as high as 1 million Chinese Yuan ($ 0.156 million), and other remuneration is also rather considerable (http://www.1000plan.org/qrjh/article/2076).

(3) Many Chinese authors with high betweenness centrality in the collaboration network also have experience of studying in the United States. Some got their Ph.D. in the USA, and some have worked as postdoctors or visiting scholar in America.

Moreover, in recent years, the Chinese government has reinforced the sponsorship for Chinese students studying abroad. For example, in a single year of 2009, China Scholarship Council (CSC) recruited a total of 12,769 for all types of state sponsored study abroad programs, among whom were 219 senior research scholars, 4,001 visiting scholars, 331 post-doctors, 2,451 for full Ph.D program, and 3,174 for joint-cultivated Ph.D program, and 5458 (about 42.7%) of them are to North America(http://www.csc.edu.cn/uploads/20101008140653820.pdf). These visiting scholars and students have more chances to build relationship with US scientists and promote collaborations in their studies, which has largely reinforced the connection between Chinese and US scientists.

## 5. Conclusion

This paper focuses on the collaboration of individual scientists between China and the USA. By analyzing the pattern of China-US scientific collaboration in nanotechnology, we find that Chinese-American scientists have been playing an important role. In the co-authorship network, Chinese-American scientists tend to have higher betweenness centrality, which means China-US collaboration in nanotechnology mainly occurs between Chinese and Chinese-American scientists. Collaboration has significant impact on scientific researches, which can be seen from the highly-cited papers. The series of polices implemented by the Chinese government to recruit oversea experts seems to contribute a lot to scientific communication, and thus facilitates China-US scientific collaboration.

## Acknowledgement

The research was supported by the project of “Specialized Research Fund for the Doctoral Program of Higher Education of China” (Grant No. 2009041110001), as well as the project of "Fundamental Research Funds for the Central Universities" (Grant No. DUT10RC(3)021).